# Robust Dirac lines against Ge vacancy and possible spin-orbit Dirac points in nonsymmorphic HfGe$_{0.92}$Te

L. Chen[1,2,††], L. Q. Zhou[1,2,††], Y. Zhou[1,2,††], C. Liu[*,3], Z. N. Guo[4], S. Y. Gao[1,2], W. H. Fan[1,2], J. F. Xu[2,3], Y. X. Guo[2,3], K, Liao[1,2], J. O. Wang[*,3], H. M. Weng[*,1,2,5], and G. Wang[*,1,2,5]

[1] *Beijing National Laboratory for Condensed Matter Physics, Institute of Physics, Chinese Academy of Sciences, Beijing 100190, China;*
[2] *University of Chinese Academy of Sciences, Beijing 100049, China*
[3] *Beijing Synchrotron Radiation Facility, Institute of High Energy Physics, Chinese Academy of Sciences, Beijing 100049, China*
[4] *Department of Chemistry, School of Chemistry and Biological Engineering, University of Science and Technology Beijing, Beijing 100083, China*
[5] *Songshan Lake Materials Laboratory, Dongguan, Guangdong 523808, China*
[††] *These authors contributed equally to this work.*
[*]Corresponding author: gangwang@iphy.ac.cn; hmweng@iphy.ac.cn; wangji@ihep.ac.cn; cliu@ihep.ac.cn

Looking for new materials with Dirac points has been a fascinating subject of research. Here we report the growth, crystal structure, and band structure of HfGe$_{0.92}$Te single crystals, featuring three different types of Dirac points. HfGe$_{0.92}$Te crystallizes in a nonsymmorphic tetragonal space group *P4/nmm* (No. 129), having square Ge-atom plane with vacancies about 8%. Despite the vacancies on Ge site, the Dirac nodal line composed of conventional Dirac points vulnerable to spin-orbit coupling (SOC) is observed using angle-resolved photoemission spectroscopy, accompanied with the robust Dirac line protected by the nonsymmorphic symmetry against both SOC and vacancies. Specially, spin-orbit Dirac points (SDPs) originated from the surface formed under SOC are hinted to exist according to our experiments and calculations. Quasi-two-dimensional (quasi-2D) characters are observed and further confirmed by angular-resolved magnetoresistance. HfGe$_{0.92}$Te is a good candidate to explore exotic topological phases or topological properties with three different types of Dirac points and a promising candidate to realize 2D SDPs.

## 1 Introduction

Exotic topological phases have aroused a lot attention especially for the potential applications of the dissipationless gapless edge states in spintronic devices [1, 2]. The discoveries of topological materials, like two-dimensional (2D) [3, 4] or three-dimensional (3D) [5] topological insulators (TIs), Chern insulators [6-8], topological crystalline insulators (TCIs) [9], and even topological semimetals such as Dirac [10-13] or Weyl semimetals [14-19], largely enhance the varieties and functionalities of materials. For example, the 3D Dirac semimetal Cd$_3$As$_2$ with Dirac point protected by crystalline rotational symmetry confirmed by angle-resolved photoemission spectroscopy (ARPES), shows giant positive magnetoresistance (MR) [20], non-trivial quantum oscillations [21], and Landau level splitting [22] in bulk crystals. By reducing the dimensionality, giant negative MR induced by chiral anomaly of separated Weyl nodes under magnetic field was observed in single $Cd_3As_2$ nanowire [23, 24]. As all of these special properties come from the Dirac points with linear dispersions near the Fermi level ($E_F$) [25], thus looking for new materials with Dirac points has been a fascinating subject of research.

*WHM*-type materials (*W* = Zr, Hf, La, *H* = Si, Ge, Sn, Sb, *M* = O, S, Se, Te) featuring *H* square nets, have been found to be a large pool of topological materials with Dirac points and linear dispersions in large energy range near $E_F$ [26, 27]. Different from the well-known Dirac semimetals $Cd_3As_2$ or $Na_3Bi$ with accidental band inversions caused by low-lying 5*s* or 6*s* atomic orbitals [10, 11], the delocalized $\pi$ bonds in the *H* square nets give rise to an essential band inversion with linear dispersions in large energy range [28]. Most of the *WHM*-type materials were proved to be novel nodal-line Dirac semimetals [29-31], some of them has linear dispersions in large energy range near $E_F$ without interference from other bulk bands [29, 30]. Once considering spin-orbit coupling (SOC), *WHM*-type materials crystalized in tetragonal phases can be considered as TCIs as a simple stacking of 2D TIs [32]. Rich exotic properties, including extremely large MR [33], electron-hole tunneling [34], and large second harmonic generation effect [35], were predicted and observed within them.

Up to now, *WHM*-type materials having Si [29-31, 34, 36], Ge [37], Sn [38], and Sb [39-44] square nets with topological properties have been extensively studied experimentally, except *WH*O (*W* = Zr, Hf, La, *H* = Si, Ge, Sn, Sb) and HfGe*M* (*M* = S, Se, Te). The monolayer of ZrSiO was proposed to be a 2D TI with a band gap up to 30 meV [26] but not experimentally realized yet for the challenge of synthesizing O-contained *WHM*-type compounds [45]. HfGe*M* (M = S, Se, Te) monolayers were predicted to host the 2D spin-orbit Dirac points (SDPs) close to $E_F$ [46]. Compared with common Dirac points which is vulnerable to SOC [47, 48], SDPs can be formed under significant SOC and intrinsically robust against SOC. It is essential to investigate SDPs to further explore the unique properties of 2D Dirac points materials, like nonlinear Hall effect [49] and so on. Though embedded with such exotic phases, the experimental research on HfGeTe is still lack for the challenge of single crystal growth.

Here we report the growth, crystal structure, and band structure of $HfGe_{0.92}Te$ single crystals with vacancies on Ge square net. The experimentally observed band structures of $HfGe_{0.92}Te$ using *in-situ* ARPES can be well described using the seven-unit-cell-thick slab of HfGeTe. The Ge vacancies only induce a little hole doping by lowering the $E_F$ about 0.2 eV, showing the robust band structure against small perturbations. Dirac-cone-like band structure with linear dispersions in large energy range ($\Delta E \sim$ 0.6 eV) was observed along $\bar{M} - \bar{\Gamma} - \bar{M}$, with the crossing points near $E_F$, which would be gapped under SOC. At high symmetry point $\bar{X}$, Dirac points at $E - E_F = -0.4$ eV protected by nonsymmorphic symmetry were observed, accompanied with anisotropic electron pockets with the cone point at $E - E_F = -0.2$ eV. For the electron pockets, a split was observed along $\bar{X} - \bar{M}$, which may hint the existence of intrinsic SDPs robust against SOC according to our calculation. The ARPES intensity map below $E_F$ was also displayed and the quasi-2D characters were observed from the non-dispersive $\bar{X} - \bar{R}$ bands and typical two-fold anisotropic angular-resolved magnetoresistance (AMR) with AMR $\propto (Bcos\theta)^2$, showing the formation of Dirac lines along $\bar{X} - \bar{R}$. Thus three types of Dirac points, the conventional Dirac points vulnerable to SOC forming Dirac nodal line, the robust Dirac points protected by the nonsymmorphic symmetry against both SOC and vacancies forming Dirac lines, and SDPs coming from the surface, were observed coexisting in $HfGe_{0.92}Te$, indicating that $HfGe_{0.92}Te$ is a good platform to explore exotic topological phases or topological properties, especially promising to realize the 2D SDPs considering its quasi-2D nature.

## 2 Materials and method

**Single Crystal Growth and Characterization.** The single crystals of HfGe$_{0.9}$Te were grown by high temperature solution method using Ge flux. The hafnium powder (Alfa Aesar, 99.5%), germanium powder (Alfa Aesar, 99.999%), and tellurium powder (Alfa Aesar, 99.9%) were mixed using a molar ratio of 2:11:1 in a fritted alumina crucible set (Canfield crucible set) [50] and sealed in a fused-silica ampoule at vacuum. The sealed ampoule was heated to 1423 K, kept for 5 h, and then slowly cooled down to 1273 K at a rate of 2 K/h. By centrifugation at 1273 K, black, shiny, and air-stable plate-like crystals as large as 2 mm × 2 mm × 0.5 mm were obtained. Single-crystal X-ray diffraction (SCXRD) was performed on a four-circle diffractometer (XtaLAB Synergy R-DW, HyPix) at 180 K with multilayer mirror graphite-monochromatized Mo $K_\alpha$ radiation ($\lambda$ = 0.71073 Å) operated at 50 kV and 40 mA. Powder X-ray diffraction (PXRD) data were collected on a PANalytical X'Pert PRO diffractometer (Cu $K_\alpha$ radiation) operated at 40 kV and 40 mA with a graphite monochromator in a reflection mode ($2\theta$ = 5°–100°, step size = 0.017°). Indexing and Rietveld refinement were performed using the DICVOL91 and FULLPROF programs [51]. Elemental analysis was conducted using a scanning electron microscope (SEM, Hitachi S-4800) equipped with an electron microprobe analyzer for semi-quantitative elemental analysis in energy-dispersive spectroscopy (EDS) mode. AMR was performed on a physical property measurement system (PPMS, Quantum Design) at 5 K. Contacts for standard four-probe configuration were established by attaching platinum wires using silver paint, resulting in a contact resistance smaller than 5 Ω. Samples for all measurements were cleaved along (00$l$) ($l$ = integer) from HfGe$_{0.92}$Te crystals using a razor blade.

**First-Principles Calculations.** We used the crystal structure of HfGeTe to simulate the electronic structures by lowing the $E_F$ about 0.2 eV in our calculations. The first-principles calculations were performed by using the Vienna *ab initio* simulation package (VASP) [52]. The generalized gradient approximation (GGA) of Perdew-Burke-Ernzerhof (PBE) type was selected to describe the exchange-correlation function [53], and the modified Becke-Johnson (mBJ) potential was used to obtain accurate bulk band results [54, 55]. In order to really simulate the surface bands of the material, we constructed the monolayer and seven-unit-cell slab model, and added both 15 Å vacuum layers to minimize the interactions between the layers. The Brillouin zone (BZ) integration was sampled by 9 × 9 × 5 k-mesh for bulk and 9 × 9 × 1 for the monolayer and seven-unit-cell slab. The cut-off energy was set to 450 eV. The atomic positions of the top and bottom unit cell for seven-unit-cell slab model were optimized. SOC was taken into account self-consistently. The tight-binding model of HfGeTe was constructed by the Wannier90 with Hf 5$d$ orbitals, Te 5$p$ orbitals, and Ge 4$p$ orbitals, which based on the maximally-localized Wannier functions (MLWF) [56]. The Fermi surface (FS) planes of HfGeTe were calculated by the WannierTools package [57].

**Electronic Structure Measurement.** Synchrotron ARPES and X-ray photoelectron spectroscopy measurements with various photon energies on HfGe$_{0.92}$Te were performed at beamline 4B9B of Beijing Synchrotron Radiation Facility (BSRF). Helium discharge lump ($h\nu$ = 21.2 eV) was also used as light source in ARPES measurement. The ARPES system is equipped with a Scienta R4000 electron analyzer and has base pressure 7 × 10$^{-11}$ Torr. The overall energy and angular resolutions were 17 meV and 0.3°, respectively. HfGe$_{0.92}$Te single crystals were cleaved *in-situ* along the (00*l*) (*l* = integer) crystal plane and measured at 7 K.

## 3 Results and Discussion

Based on the results of SCXRD (Table a1, a2), HfGe$_{0.92}$Te crystallizes in a nonsymmorphic tetragonal space group *P*4/*nmm* (No. 129) with lattice parameters $a = b = 3.8467(3)$ Å, $c = 8.4551(16)$ Å, and $\alpha = \beta = \gamma = 90°$ (Table a1), featuring square Ge-atom plane sandwiched by two HfTe layers (Figure 1(a)). As a member of *WHM* family, its structure is similar to the well-known "111"-type iron-based superconductor LiFeAs [58, 59] and isostructural to nodal-line Dirac semimetal ZrSiS [33]. By refining the site occupancies, it was found there are about 8% vacancies in the Ge sublattice, being consistent with the result of EDS (Figure a1). Considering the random distribution of Ge vacancies, HfGe$_{0.92}$Te should still has the same symmetries as those of HfGeTe, including one inversion symmetry, a four-fold axis, three mirror symmetries, and especially the nonsymmorphic *n*-glide mirror perpendicular to *c* axis. Only (00*l*) (*l* = integer) diffraction peaks are observed in the PXRD pattern, indicating that the plate surface is perpendicular to *c* axis (Figure 1(b)).

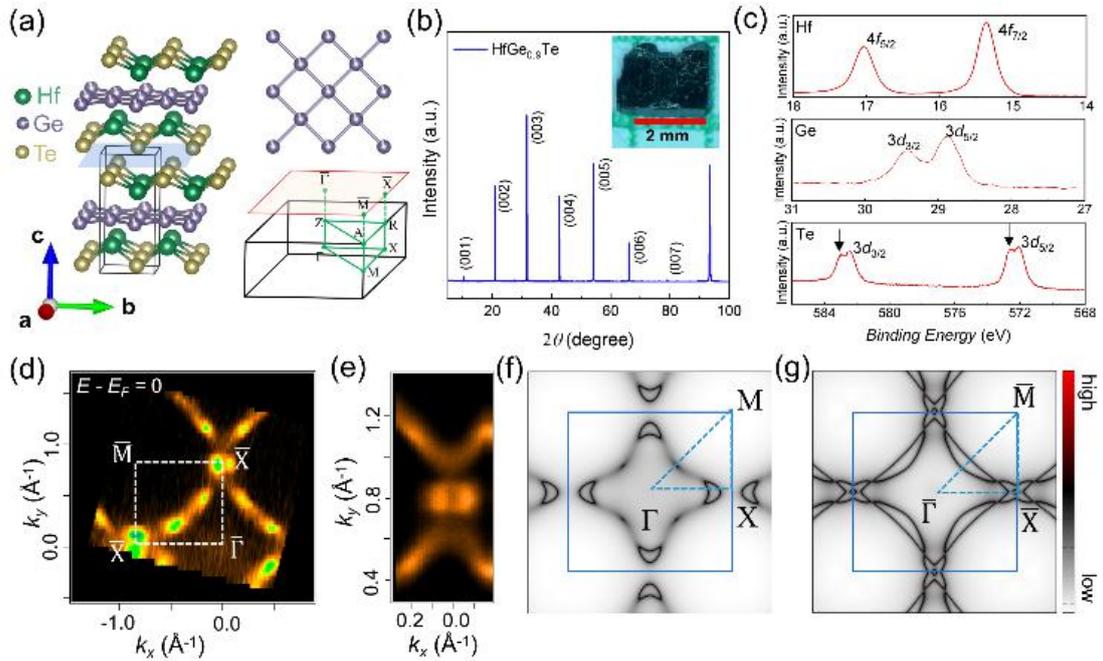

**Figure 1** (Color online) Crystal structure and FS of HfGe$_{0.92}$Te. (a) Crystal structure of HfGe$_{0.92}$Te determined by SCXRD and the BZ of bulk (black) and corresponding surface (red). The light blue plane indicates the cleavage plane perpendicular to *c* axis. The black cuboid demonstrates the unit cell and the monolayer used in the calculation. (b) PXRD pattern of HfGe$_{0.92}$Te single crystal showing (00*l*) (*l* = integer) diffraction peaks. The inset is the optical photograph of the single crystal. (c) Core-level spectra of Hf 4*f*, Ge 3*d*, and Te 3*d* after cleavage, respectively. The arrows donate the split of corresponding peaks. (d) FS intensity plot in $\overline{\Gamma} - \overline{M} - \overline{X}$ plane of HfGe$_{0.92}$Te recorded at $hv = 46$ eV, obtained by integrating the spectral weight within ±10 meV with respect to $E_F$. (e) Zoomed FS intensity plot around $\overline{X}$. Calculated FS for HfGeTe (f) bulk crystal and (g) monolayer.

As shown in Figure 1(c), the *in-situ* core-level spectra after cleavage show characteristic peaks of Hf, Ge, and Te atoms, indicating a clean (001) surface of HfGe$_{0.92}$Te (Figure a2). Both peaks of Te ($3d_{1/2}$ and $3d_{3/2}$) are split, indicating that there may exist two chemical environments of Te ions in the terminated layer. This may suggest the emerging Te-Hf terminal after cleavage (Figure 1(a)), which is consistent with the relatively weak coupling between two adjacent HfTe layers, similar to the S-Zr terminal in ZrSiS [35]. For the measured FS near $E_F$, a diamond-shaped FS formed by the Dirac points along $\bar{\Gamma}-\bar{M}$ direction can be observed (Figure 1(d)), similar to the diamond-shaped dispersions observed in ZrSiS [29]. Additionally, tiny anisotropic electron pockets exist around $\bar{X}$ (Figure 1(e)), which is distinctly different from the calculations of bulk HfGeTe with no pockets at X (Figure 1(f)). We note that electron pockets at $\bar{X}$ were predicted to exist in the monolayer of HfGeTe [46], and Figure 1(g) shows the FS with electron pockets at $\bar{X}$ in $\bar{\Gamma}-\bar{M}-\bar{X}$ plane of HfGeTe monolayer according to our calculation. This may suggest that the observed anisotropic electron pockets come from the surface of HfGe$_{0.92}$Te, which has been observed previously in HfSiS [36] and ZrSnTe [38].

Figure 2(a) shows the calculated band dispersions along the high-symmetry lines for HfGeTe bulk crystal. Like previously reported in most *WHM-type* compounds [26], HfGeTe exhibits two kinds of Dirac points: Dirac points on Γ-M (blue circles) which form a diamond-shaped Dirac node line and four-fold degenerated Dirac points (purple circles) at high symmetry points of BZ like X and R. When considering the SOC effect, the Dirac node line is gapped, whereas the four-fold degenerated Dirac points protected by the nonsymmorphic glide planes could survive. Compared with the calculated band dispersions of the bulk crystal, the electron bands at $\bar{X}$ for the monolayer and seven-unit-cell-thick slab are shifted down by ~ 1 eV, forming electron pockets centered at $\bar{X}$ (Figure 2(b) and (c)). These pockets mainly come from the Hf 5*d* orbitals, which would give rise to an exotic SDP (green circle) when considering the large Rashba SOC effect [46]. For the global gap induced by SOC, the bulk crystal and monolayer of HfGeTe could be classified into a TCI with $Z_4 = 1$ [32] and a 2D TI with $Z_2 = 1$ [26], respectively.

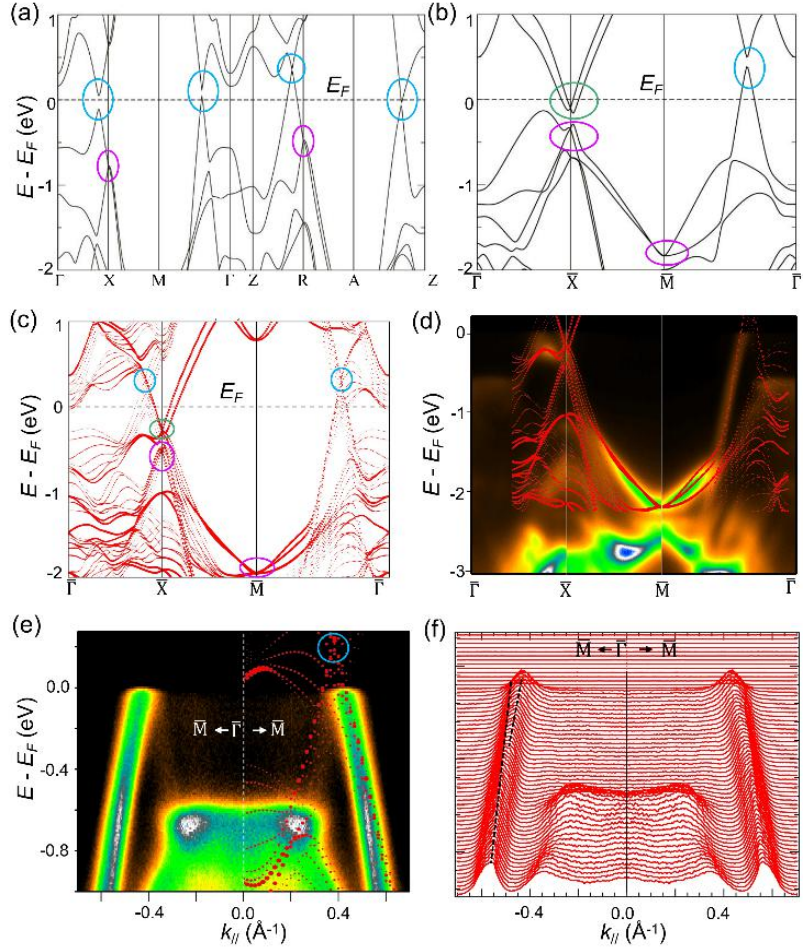

**Figure 2** (Color online) Electronic structure of HfGe$_{0.92}$Te. (a)-(c) Calculated band dispersions along the high-symmetry lines with SOC for HfGeTe bulk crystal, monolayer, and a seven-unit-cell-thick slab, respectively. Blue, purple, and green circles highlight three kinds of Dirac points. (d) Experimentally observed band structures for HfGe$_{0.92}$Te along the high-symmetry lines $\bar{\Gamma} - \bar{X} - \bar{M} - \bar{\Gamma}$ at $h\nu = 21$ eV. (e) Band structures along $\bar{M} - \bar{\Gamma} - \bar{M}$ and (f) corresponding MDC. The red dotted lines in (d) and (e) are extracted from the calculation in (c) by lowering the $E_F$ about 0.2 eV. The black dotted lines in (f) show the multiple bands feature.

The experimentally observed band structures can be well described using the seven-unit-cell-thick slab of HfGeTe (Figure 2(d)), which is rational considering that the signals of escaped photoelectrons mainly come from the several top unit cells close to the terminal in ARPES experiments. The clear difference between the experimentally observed band structures for HfGe$_{0.92}$Te and calculated band dispersions of seven-unit-cell-thick slab of HfGeTe is the location of $E_F$, as the Ge vacancies would cause a ~ 0.2 eV hole doping. Figure 2(e) shows the band structure along $\bar{M} - \bar{\Gamma} - \bar{M}$, exhibiting a Dirac-like band structure with crossing points (blue circle) a little bit higher than $E_F$ and linear dispersions in large energy range ($\Delta E \sim 0.6$ eV). As described above, this kind of Dirac points would be gapped under SOC (Figure 2(a)). In the momentum distribution curve (MDC) (Figure 2(f)), the feature of multiple bands can be observed and attributed to the parallel bands of the seven-unit-cell-thick slab according to the calculation, similar to what happened in ZrSnTe [38].

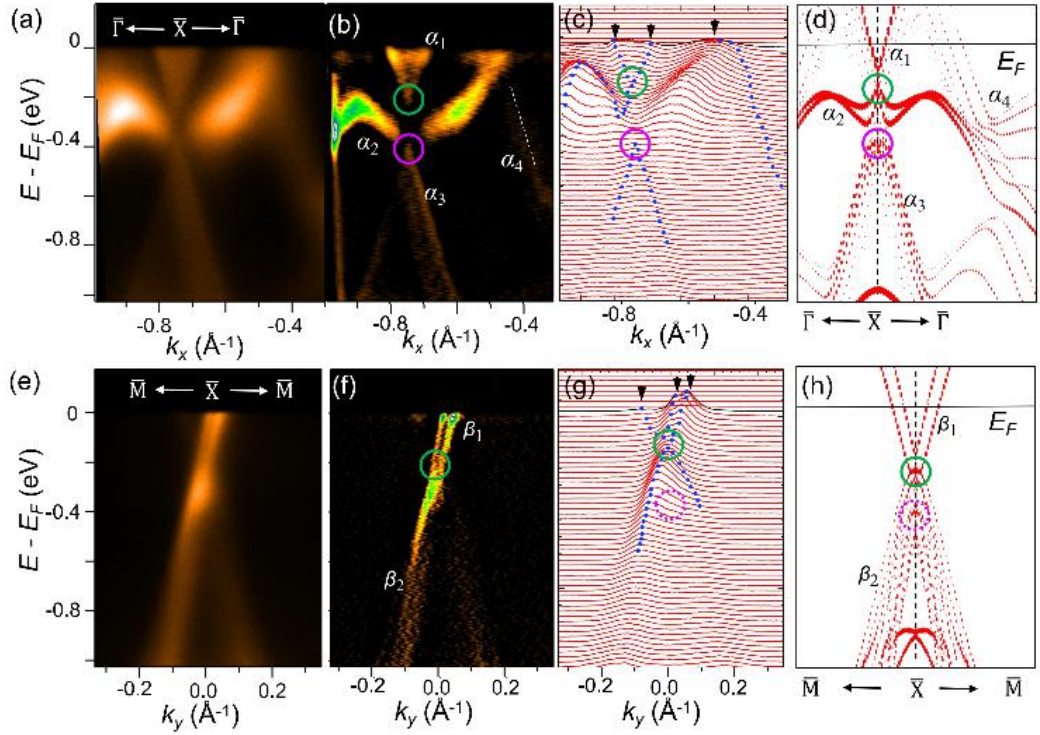

**Figure 3** (Color online) Band structure for HfGe$_{0.92}$Te around $\bar{X}$. (a) Photoemission intensity plot along $\bar{\Gamma} - \bar{X} - \bar{\Gamma}$ with $hv$ = 21.2 eV. (b) Second derivative intensity plot and (c) MDC plot of (a). (d) Calculated band structure along $\bar{\Gamma} - \bar{X} - \bar{\Gamma}$ for a seven-unit-cell-thick slab. The intensity of the red color scales the spectral weight projected to the top unit cells. (e)–(h) The same as (a)–(d) but along $\bar{M} - \bar{X} - \bar{M}$. Purple and green circles highlight two kinds of Dirac points. The arrows donate the crossings of the dispersion with $E_F$ (black line) and dotted lines outline the shape of the dispersions.

The band structures for HfGe$_{0.92}$Te along $\bar{\Gamma} - \bar{X} - \bar{\Gamma}$ and $\bar{M} - \bar{X} - \bar{M}$ are summarized in Figure 3. Dirac-cone-shaped band structure with anisotropic electron pockets around $\bar{X}$ can be clearly observed (Figure 3(c) and (g)), which is consistent with the calculation of the seven-unit-cell-thick slab (Figure 3(d) and (h)). Along $\bar{\Gamma} - \bar{X} - \bar{\Gamma}$, four sets of dispersions with different intensity can be observed, resulting in different types of Dirac points. The dispersions with the highest intensity ($\alpha_2$) and the downward linear dispersions ($\alpha_3$) along $\bar{X} - \bar{\Gamma}$ form Dirac points (purple circles) at $\bar{X}$ around $E - E_F$ = -0.4 eV. According to the calculation and symmetry analysis, the Dirac points at $\bar{X}$ are protected by nonsymmorphic glide planes and robust against perturbations, specifically the Ge vacancies in the case of HfGe$_{0.92}$Te. The small electron pockets ($\alpha_1$) crossing $E_F$ with the cone points (green circles) at $E - E_F$ = -0.2 eV can be clearly observed, which is quite different from the calculation of bulk crystals. The linear dispersions with the lowest intensity ($\alpha_4$) cross the $E_F$, which would induce the formation of Dirac points above $E_F$ (blue circle in Figure 2(c)). This kind of Dirac points should form Dirac lines combined with the Dirac points along $\bar{M} - \bar{\Gamma} - \bar{M}$, but vulnerable to SOC. Along $\bar{M} - \bar{X} - \bar{M}$, the Dirac-cone-shaped band structure should be resolved to the electron pockets with higher intensity ($\beta_1$) and the Dirac points with lower intensity ($\beta_2$) according to the calculation (Figure 3(h)). Because of the possible matrix element effect of ARPES measurement, the Dirac points located at $E - E_F$ = -0.4 eV are not that clear as observed along $\bar{\Gamma} - \bar{X} - \bar{\Gamma}$ (Figure a3).

As predicted in HfGeTe monolayer, a 2D SDP robust against both the SOC and vacancies can be formed for the split caused by Rashba SOC effect of Hf, and the location of the 2D SDP can be tuned by uniaxial strain and biaxial strain or vacancies [46]. For the electron pockets, a clear split is observed in the second derivative intensity and MDC plot (Figure 3(f) and (g)). The split can be regarded as the signal of the existence of SDP in HfGe$_{0.92}$Te and its location ($E - E_F$ = -0.2 eV) is closer to $E_F$ than that of HfSiS ($E - E_F$ = -0.4 eV) [36], which has more potential to embed exotic quantum transport properties or applications [46, 49]. This SDP is protected by the nonsymmorphic glide plane and time-reversal symmetry in HfGe$_{0.92}$Te, resulting in the degeneracy robust against SOC [46]. It should be pointed out that the split has only been observed in HfSiS [36], not in ZrSiS [29, 30] or ZrSnTe [38], which may be attributed to the larger SOC effect of Hf with larger atomic mass (Z = 72) than that of Zr (Z = 40).

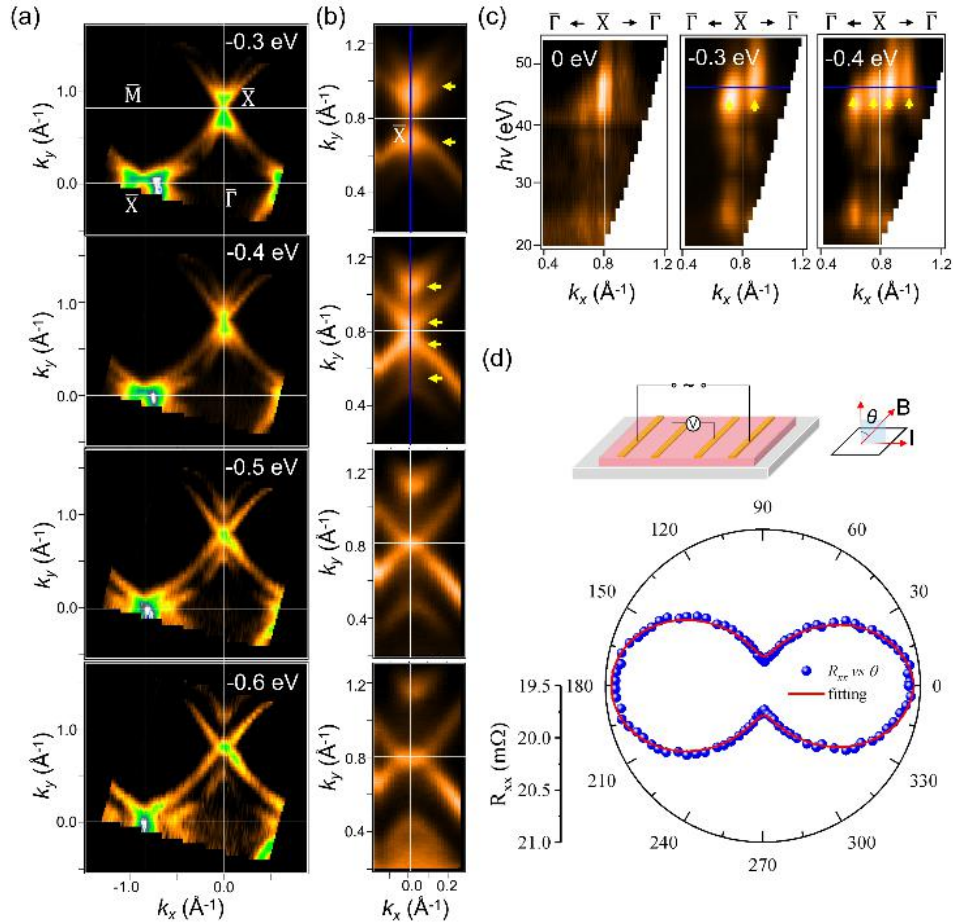

**Figure 4** (Color online) 3D band structure and quasi-2D FS of HfGe$_{0.92}$Te. (a) ARPES intensity maps of $\bar{\Gamma} - \bar{M} - \bar{X}$ plane at $E - E_F$ = -0.3 eV, -0.4 eV, -0.5 eV, and -0.6 eV with $hv$ = 46 eV. (b) Zoomed ARPES intensity maps of $\bar{\Gamma} - \bar{M} - \bar{X}$ plane around $\bar{X}$. (c) ARPES intensity maps along $\bar{X} - \bar{R}$ for $E - E_F$ = 0 eV, -0.3 eV, and -0.4 eV measured with varying $hv$ from 20 eV to 54 eV. The blue lines and yellow arrows donate the cut positions and corresponding dispersions in (b) and (c), respectively. (d) Schematic of AMR experimental configuration and polar plot at 9 T for HfGe$_{0.92}$Te.

To more precisely observe the electronic structure of HfGe$_{0.92}$Te near the $E_F$, the ARPES intensity maps of $\bar{\Gamma} - \bar{M} - \bar{X}$ plane at $E - E_F$ = -0.3 eV, -0.4 eV, -0.5 eV, and -0.6 eV were measured with $hv$ = 46 eV (Figure 4(a)). Compared with the FS of HfGe$_{0.92}$Te (Figure 1(e)), the electron pockets vanish absolutely at $E - E_F$ = -0.3 eV, indicating that the SDPs locate quiet close to $E_F$. By further increasing the binding energy from $E - E_F$ = -0.3 eV to $E - E_F$ = -0.6 eV, a banana-like feature emerges and gradually expands along $\bar{\Gamma} - \bar{M}$, which finally evolves into two diamonds at $E - E_F$ = -0.6 eV. In addition, small anisotropic hole pockets emerge and expend at $\bar{X}$ (Figure 4(b)), indicating the location of Dirac points protected by nonsymmorphic symmetry. These results were repeatable in the ARPES intensity maps of $\bar{\Gamma} - \bar{M} - \bar{X}$ plane measured with $hv$ = 21.2 eV (Figure a4).

The band structure along $\bar{X} - \bar{R}$ has been measured by varying the photon energy from 20 eV to 54 eV. As shown in Figure 4(c), only the change of intensity is observed along $\bar{X} - \bar{R}$ for $E - E_F$ > -1 eV, whereas the dispersions along $\bar{X} - \bar{R}$ are observed for $E - E_F$ < -1 eV (Figure a5), showing a quasi-2D FS. The 2D-like feature of FS in HfGe$_{0.92}$Te is consistent with its reduced interlayer couplings compared with those of ZrSiS [26, 31, 60] and is further supported by our measurement of AMR. With the current being applied in $ab$ plane and the field being rotated in a manner shown in Figure 4(d), the AMR of HfGe$_{0.92}$Te exhibits a typical two-fold anisotropy and classical Lorentz-type MR with AMR $\propto (Bcos\theta)^2$. These behaviors are expected to happen in materials with 2D/quasi-2D electronic structure, such as ZrSiTe [31], whereas a "butterfly-shaped" AMR would be observed in materials with a 3D-like FS, such as ZrSiS [47, 61] or ZrSiSe [31]. Considering the 2D/quasi-2D characters of HfGe$_{0.92}$Te, the Dirac points at $\bar{X}$ protected by the nonsymmorphic symmetry would form Dirac lines along $\bar{X} - \bar{R}$ against both SOC and vacancies. Moreover, the 2D/quasi-2D characters usually correspond to a small exfoliation energy, making HfGe$_{0.92}$Te a potential candidate to realize the 2D SDPs by mechanical exfoliation from the bulk crystals.

## 4 Conclusions

In conclusion, we successfully grown the single crystals of HfGe$_{0.92}$Te, a member of *WHM* family but with vacancies on its Ge square net. The ARPES-observed band structures of HfGe$_{0.92}$Te can be well described using the seven-unit-cell-thick slab of HfGeTe, whereas the Ge vacancies only induce a little hole doping and lower $E_F$ about 0.2 eV. The Dirac-cone-like band structure with linear dispersions in large energy range ($\Delta E \sim 0.6$ eV) was observed along $\bar{\Gamma} - \bar{M}$ with its crossing points near $E_F$, which would be gapped under SOC. At high symmetry point $\bar{X}$, Dirac points protected by nonsymmorphic symmetry and anisotropic electron pockets were observed. For the electron pockets, a split was observed along $\bar{X} - \bar{M}$, which hint the existence of SDPs under SOC according to our calculation. Quasi-2D characters arise from the non-dispersive $\bar{X} - \bar{R}$ bands and typical two-fold anisotropic AMR with AMR $\propto (Bcos\theta)^2$, showing the formation of Dirac lines along $\bar{X} - \bar{R}$. Thus three types of Dirac points, the conventional Dirac points vulnerable to SOC forming Dirac nodal line, the robust Dirac points protected by the nonsymmorphic symmetry against both SOC and vacancies forming Dirac lines along $\bar{X} - \bar{R}$, and possible SDPs coming from the surface, coexist in HfGe$_{0.92}$, indicating that HfGe$_{0.92}$Te is a good candidate to explore exotic topological phases or topological properties, especially promising to realize the 2D SDPs considering its quasi-2D nature.

# Reference

# Acknowledgement

L. Chen and G. Wang would like to thank Prof. X. L. Chen and Prof. T. Qian, Institute of Physics, Chinese Academy of Sciences, Prof. J. Ma of Shanghai Jiao Tong University for useful discussions. This work was partially supported by the National Natural Science Foundation of China (51832010, 51902055, 11925408, 12005251, and 11921004) and the National Key Research and Development Program of China (2018YFE0202600, 2018YFA0305700, and 2017YFA0302902).

# Reference

1. M. Z. Hasan, C. L. Kane, Rev. Mod. Phys. **82**, 3045 (2010).
2. X. L. Qi, S. C. Zhang, Rev. Mod. Phys. **83**, 1057 (2011).
3. B. A. Bernevig, T. L. Hughes, S. C. Zhang, Science **314**, 1757 (2006).
4. M. Konig, S. Wiedmann, C. Brune, A. Roth, H. Buhmann, L. W. Molenkamp, X. L. Qi, S. C. Zhang, Science **318**, 766 (2007).
5. Y. L. Chen, J. G. Analytis, J. H. Chu, Z. K. Liu, S. K. Mo, X. L. Qi, H. J. Zhang, D. H. Lu, X. Dai, Z. Fang, S. C. Zhang, I. R. Fisher, Z. Hussain, Z. X. Shen, Science **325**, 178 (2009).
6. R. Yu, W. Zhang, H. J. Zhang, S. C. Zhang, X. Dai, Z. Fang, Science **329**, 61 (2010).
7. C. Z. Chang, J. S. Zhang, X. Feng, J. Shen, Z. C. Zhang, M. H. Guo, K. Li, Y. B. Ou, P. Wei, L. L. Wang, Z. Q. Ji, Y. Feng, S. H. Ji, X. Chen, J. F. Jia, X. Dai, Z. Fang, S. C. Zhang, K. He, Y. Y. Wang, L. Lu, X. C. Ma, Q. K. Xue, Science **340**, 167 (2013).
8. H. M. Weng, R. Yu, X. Hu, X. Dai, Z. Fang, Advances in Physics **64**, 227 (2015).
9. L. Fu, Phys. Rev. Lett. **106**, 106802 (2011).
10. Z. J. Wang, Y. Sun, X. Q. Chen, C. Franchini, G. Xu, H. M. Weng, X. Dai, Z. Fang, Phys. Rev. B **85**, 195320 (2012).
11. Z. J. Wang, H. M. Weng, Q. S. Wu, X. Dai, Z. Fang, Phys. Rev. B **88**, 125427 (2013).
12. Z. K. Liu, B. Zhou, Y. Zhang, Z. J. Wang, H. M. Weng, D. Prabhakaran, S. K. Mo, Z. X. Shen, Z. Fang, X. Dai, Z. Hussain, Y. L. Chen, Science **343**, 864 (2014).
13. Z. K. Liu, J. Jiang, B. Zhou, Z. J. Wang, Y. Zhang, H. M. Weng, D. Prabhakaran, S. K. Mo, H. Peng, P. Dudin, T. Kim, M. Hoesch, Z. Fang, X. Dai, Z. X. Shen, D. L. Feng, Z. Hussain, Y. L. Chen, Nat. Mater. **13**, 677 (2014).
14. H. M. Weng, C. Fang, Z. Fang, B. A. Bernevig, X. Dai, Phys. Rev. X **5**, 011029 (2015).
15. B. Q. Lv, H. M. Weng, B. B. Fu, X. P. Wang, H. Miao, J. Ma, P. Richard, X. C. Huang, L. X. Zhao, G. F. Chen, Z. Fang, X. Dai, T. Qian, H. Ding, Phys. Rev. X **5**, 031013 (2015).
16. S. M. Huang, S. Y. Xu, I. Belopolski, C. C. Lee, G. Q. Chang, B. K. Wang, N. Alidoust, G. Bian, M. Neupane, C. L. Zhang, S. Jia, A. Bansil, H. Lin, M. Z. Hasan, Nat. Commun. **6**, 7373 (2015).
17. S. Y. Xu, I. Belopolski, N. Alidoust, M. Neupane, G. Bian, C. L. Zhang, R. Sankar, G. Q. Chang, Z. J. Yuan, C. C. Lee, S. M. Huang, H. Zheng, J. Ma, D. S. Sanchez, B. K. Wang, A. Bansil, F. C. Chou, P. P. Shibayev, H. Lin, S. Jia, M. Z. Hasan, Science **349**, 613 (2015).
18. B. Q. Lv, N. Xu, H. M. Weng, J. Z. Ma, P. Richard, X. C. Huang, L. X. Zhao, G. F. Chen, C. E. Matt, F. Bisti, V. N. Strocov, J. Mesot, Z. Fang, X. Dai, T. Qian, M. Shi, H. Ding, Nat. Phys. **11**, 724 (2015).
19. S. Y. Xu, N. Alidoust, I. Belopolski, Z. J. Yuan, G. Bian, T. R. Chang, H. Zheng, V. N. Strocov, D. S. Sanchez, G. Q. Chang, C. L. Zhang, D. X. Mou, Y. Wu, L. N. Huang, C. C. Lee, S. M. Huang, B. K. Wang, A. Bansil, H. T. Jeng, T. Neupert, A. Kaminski, H. Lin, S. Jia, M. Z. Hasan, Nat. Phys. **11**, 748

(2015).

20. T. Liang, Q. Gibson, M. N. Ali, M. H. Liu, R. J. Cava, N. P. Ong, Nat. Mater. **14**, 280 (2015).
21. L. P. He, X. C. Hong, J. K. Dong, J. Pan, Z. Zhang, J. Zhang, S.Y. Li, Phys. Rev. Lett. **113**, 246402 (2014).
22. S. Jeon, B. B. Zhou, A. Gyenis, B. E. Feldman, I. Kimchi, A. C. Potter, Q. D. Gibson, R. J. Cava, A. Vishwanath, A. Yazdani, Nat. Mater. **13**, 851 (2014).
23. C. Z. Li, L. X. Wang, H. W. Liu, J. Wang, Z. M. Liao, D. P. Yu, Nat. Commun. **6**, 10137 (2015).
24. E. V. Gorbar, V. A. Miransky, I. A. Shovkovy, Phys. Rev. B **88**, 165105 (2013).
25. J. Y. Feng, Y. Pang, D. S. Wu, Z. J. Wang, H. M. Weng, J. Q. Li, X. Dai, Z. Fang, Y. G. Shi, L. Lu, Phys. Rev. B **92**, 081306 (2015).
26. Q. N. Xu, Z. D. Song, S. M. Nie, H. M. Weng, Z. Fang, X. Dai, Phys. Rev. B **92**, 205310 (2015).
27. I. Lee, S. I. Hyun, J. H. Shim, Phys. Rev. B **103**, 165106 (2021).
28. W. Tremel, R. Hoffmann, J. Am. Chem. Soc. **109**, 124 (1987).
29. L. M. Schoop, M. N. Ali, C. Strasser, A. Topp, A. Varykhalov, D. Marchenko, V. Duppel, S. S. P. Parkin, B. V. Lotsch, C. R. Ast, Nat. Commun. **7**, 11696 (2016).
30. M. Neupane, I. Belopolski, M. M. Hosen, D. S. Sanchez, R. Sankar, M. Szlawska, S. Y. Xu, K. Dimitri, N. Dhakal, P. Maldonado, P. M. Oppeneer, D. Kaczorowski, F. C. Chou, M. Z. Hasan, T. Durakiewicz, Phys. Rev. B **93**, 201104 (2016).
31. J. Hu, Z. J. Tang, J. Y. Liu, X. Liu, Y. L. Zhu, D. Graf, K. Myhro, S. Tran, C. N. Lau, J. Wei, Z. Q. Mao, Phys. Rev. Lett. **117**, 016602 (2016).
32. T. T. Zhang, Y. Jiang, Z. D. Song, H. Huang, Y. Q. He, Z. Fang, H. M. Weng, C. Fang, Nature **566**, 475 (2019).
33. Y. Y. Lv, B. B. Zhang, X. Li, S. H. Yao, Y. B. Chen, J. Zhou, S. T. Zhang, M. H. Lu, Y. F. Chen, Appl. Phys. Lett. **108**, 244101 (2016).
34. M. R. van Delft, S. Pezzini, T. Khouri, C. S. A. Muller, M. Breitkreiz, L. M. Schoop, A. Carrington, N. E. Hussey, S. Wiedmann, Phys. Rev. Lett. **121**, 256602 (2018).
35. S. M. Chi, F. Liang, H. X. Chen, W. D. Tian, H. Zhang, H. H. Yu, G. Wang, Z. S. Lin, J. P. Hu, H. J. Zhang, Adv. Mater. (Weinheim, Ger.), **32**, 1904498 (2020).
36. D. Takane, Z. W. Wang, S. Souma, K. Nakayama, C. X. Trang, T. Sato, T. Takahashi, Y. Ando, Phys. Rev. B **94**, 121108 (2016).
37. M. M. Hosen, K. Dimitri, A. Aperis, P. Maldonado, I. Belopolski, G. Dhakal, F. Kabir, C. Sims, M. Z. Hasan, D. Kaczorowski, T. Durakiewicz, P. M. Oppeneer, M. Neupane, Phys. Rev. B **97**, 121103 (2018).
38. R. Lou, J. Z. Ma, Q. N. Xu, B. B. Fu, L. Y. Kong, Y. G. Shi, P. Richard, H. M. Weng, Z. Fang, S. S. Sun, Q. Wang, H. C. Lei, T. Qian, H. Ding, S. C. Wang, Phys. Rev. B **93**, 241104(R) (2016).
39. R. Singha, A. Pariari, B. Satpati, P. Mandal, Phys. Rev. B **96**, 245138 (2017).
40. M. M. Hosen, G. Dhakal, K. Dimitri, P. Maldonado, A. Aperis, F. Kabir, C. Sims, P. Riseborough, P. M. Oppeneer, D. Kaczorowski, T. Durakiewicz, M. Neupane, Sci. Rep. **8**, 13283 (2018).
41. L. M. Schoop, A. Topp, J. Lippmann, F. Orlandi, L. Muchler, M. G. Vergniory, Y. Sun, A. W. Rost, V. Duppel, M. Krivenkov, S. Sheoran, P. Manuel, A. Varykhalov, B. H. Yan, R. K. Kremer, C. R. Ast, B. V. Lotsch, Sci. Adv. **4**, eaar2317 (2018).
42. Y. Qian, Z. Tan, T. Zhang, J. Gao, Z. Wang, Z. Fang, C. Fang, H. Weng, Sci. China-Phys. Mech. Astron. **63**, 107011 (2020).
43. S. S. Yue, Y. T. Qian, M. Yang, D. Y. Geng, C. J. Yi, S. Kumar, K. Shimada, P. Cheng, L. Chen, Z. J. Wang, H. M. Weng, Y. G. Shi, K. H. Wu, B. J. Feng, Phys. Rev. B **102**, 155109 (2020).
44. Y. Wang, Y. T. Qian, M. Yang, H. X. Chen, C. Li, Z. Y. Tan, Y. Q. Cai, W. J. Zhao, S. Y. Gao, Y. Feng,

S. Kumar, E. F. Schwier, L. Zhao, H. M. Weng, Y. G. Shi, G. Wang, Y. T. Song, Y. B. Huang, K. Shimada, Z. Y. Xu, X. J. Zhou, G. D. Liu, Phys. Rev. B **103**, 125131 (2021).

45 H. Onken, K. Vierheilig, H. Hahn, Z Anorg Allg Chem **333**, 267 (1964).
46 S. Guan, Y. Liu, Z. M. Yu, S. S. Wang, Y. G. Yao, S. Y. A. Yang, Phys. Rev. Mater. **1**, 054003 (2017).
47 X. F. Wang, X. C. Pan, M. Gao, J. H. Yu, J. Jiang, J. R. Zhang, H. K. Zuo, M. H. Zhang, Z. X. Wei, W. Niu, Z. C. Xia, X. G. Wan, Y. L. Chen, F. Q. Song, Y. B. Xu, B. G. Wang, G. H. Wang, R. Zhang, Adv. Electron. Mater. **2**, 1600228 (2016).
48 Y. Yao, F. Ye, X. L. Qi, S. C. Zhang, Z. Fang, Phys. Rev. B **75**, 041401 (2007).
49 Y. J. Jin, B. B. Zheng, X. L. Xiao, Z. J. Chen, Y. Xu, H. Xu, Phys. Rev. Lett. **125**, 116402 (2020).
50 P. C. Canfield, T. Kong, U. S. Kaluarachchi, N. H. Jo, Philos. Mag., **96**, 84 (2016).
51 J. Rodríguez-Carvajal, FullProf (CEA/Saclay, France, 2001).
52 G. Kresse, J. Furthmuller, Phys. Rev. B **54**, 11169 (1996).
53 J. P. Perdew, K. Burke, M. Ernzerhof, Phys. Rev. Lett. **78**, 1396 (1997).
54 F. Tran, P. Blaha, Phys. Rev. Lett. **102**, 226401 (2009).
55 A. D. Becke, E. R. Johnson, J. Chem. Phys. **124**, (2006).
56 A. A. Mostofi, J. R. Yates, G. Pizzi, Y. S. Lee, I. Souza, D. Vanderbilt, N. Marzari, Comput. Phys. Commun. **185**, 2309 (2014).
57 Q. S. Wu, S. N. Zhang, H. F. Song, M. Troyer, A. A. Soluyanov, Comput. Phys. Commun. **224**, 405 (2018).
58 X. C. Wang, Q. Q. Liu, Y. X. Lv, W. B. Gao, L. X. Yang, R. C. Yu, F. Y. Li, C. Q. Jin, Solid State Commun. **148**, 538 (2008).
59 J. H. Tapp, Z. J. Tang, B. Lv, K. Sasmal, B. Lorenz, P. C. W. Chu, A. Phys. Rev. B **78**, 060505(R) (2008).
60 C. C. Wang, T. Hughbanks, Inorg. Chem. **34**, 5524 (1995).
61 M. N. Ali, L. M. Schoop, C. Garg, J. M. Lippmann, E. Lara, B. Lotsch, S. S. P. Parkin, Sci. Adv. **2**, e1601742 (2016).

**Supporting Information**

**Appendix A1 Crystal structure of HfGe$_{0.9}$Te**

**Table a1.** Crystallographic data from the Rietveld refinement for HfGe$_{0.92}$Te.

| Empirical formula | HfGe$_{0.92}$Te |
|---|---|
| Formula weight | 372.875 g/mol |
| Space group / Z | $P4/nmm$ (No. 129) /2 |
| Unit cell dimensions | $a$ = 3.8467(3) Å  $\alpha$ = 90° |
|  | $b$ = 3.8467 (3) Å  $\beta$ = 90° |
|  | $c$ = 8.4551 (16) Å  $\gamma$ = 90° |
| Volume / $d_{cal}$ | 125.11(3) Å$^3$ / 9.898 g/cm$^3$ |
| Reflections collected/R(int) | 331 / 0.0548 |
| Data / restraints / parameters | 98 / 0 / 10 |
| Goodness-of-fit on $F^2$ | 1.075 |
| Final R indices [I > 2sigma(I)] | $R_1$ = 0.0522, $wR_2$ = 0.1418 |
| R indices (all data) | $R_1$ = 0.0568, $wR_2$ = 0.1455 |
| Largest diff. peak and hole | 3.828 and -3.097 e.Å$^{-3}$ |

**Table a2.** Atomic coordinates and equivalent isotropic displacement parameters for HfGe$_{0.92}$Te.

| Atom | Wyck. | Site symm. | x/a | y/b | z/c | Occ. | U(eq)(Å$^2$) |
|---|---|---|---|---|---|---|---|
| Hf | 2c | 4mm | 0.7500 | 0.7500 | 0.7479(3) | 1.000 | 0.0131(10) |
| Te | 2c | 4mm | 0.2500 | 0.2500 | 0.8788(5) | 1.000 | 0.0136(12) |
| Ge | 2b | -4m2 | 0.7500 | 0.2500 | 0.5000 | 0.92(3) | 0.010(2) |

**Appendix A2 *In-situ* angle-resolved photoemission spectroscopy measurement**

Figure a1 shows the energy dispersive spectra (EDS) of HfGe$_{0.92}$Te and the *in-situ* low energy electron diffraction (LEED) pattern after cleavage. The mole ratio of Hf : Ge : Te is nearly 1.03(3) : 0.87(3) : 1.00, which may suggest vacancies on Ge site. Considering both EDS and SCXRD, the chemical

composition is decided to be HfGe$_{0.92}$Te. Typical $C_4$ in-plane symmetry can be clearly observed in the LEED pattern.

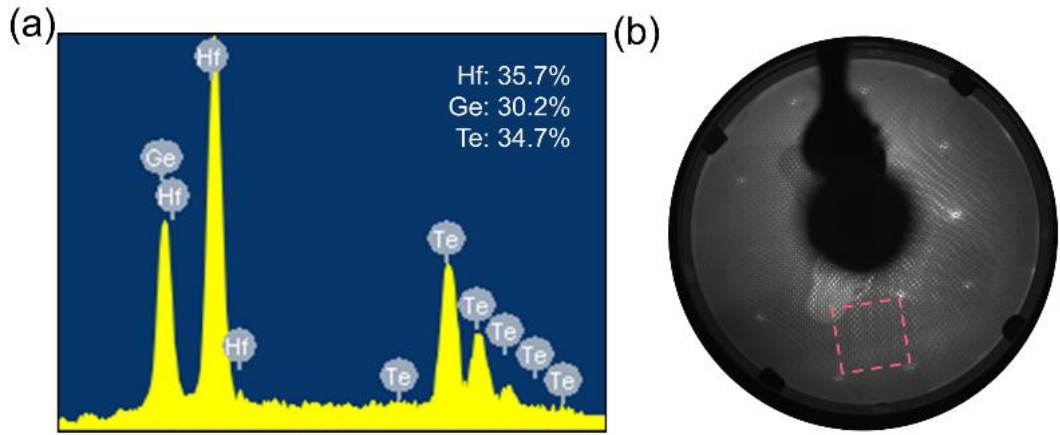

**Figure a1.** (a) EDS of HfGe$_{0.92}$Te. (b) LEED pattern of HfGe$_{0.92}$Te at $hv$ = 98 eV after cleavage.

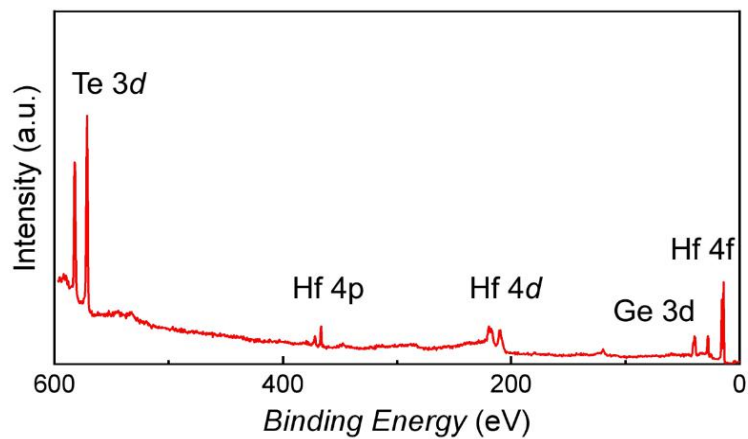

**Figure a2.** Core-level spectrum of HfGe$_{0.92}$Te after cleavage recorded at a photon energy of $hv$ = 700 eV. No peaks of C and O are observed, indicating the split of Te 3$d$ peaks is an intrinsic feature.

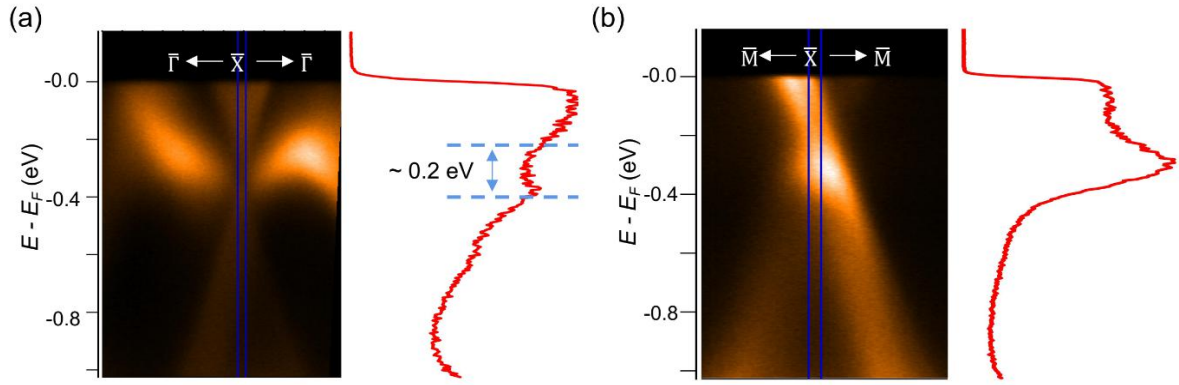

**Figure a3.** Energy distribution cures (EDCs) at $\bar{X}$ for (a) $\bar{\Gamma} - \bar{X} - \bar{\Gamma}$ and (b) $\bar{M} - \bar{X} - \bar{M}$.

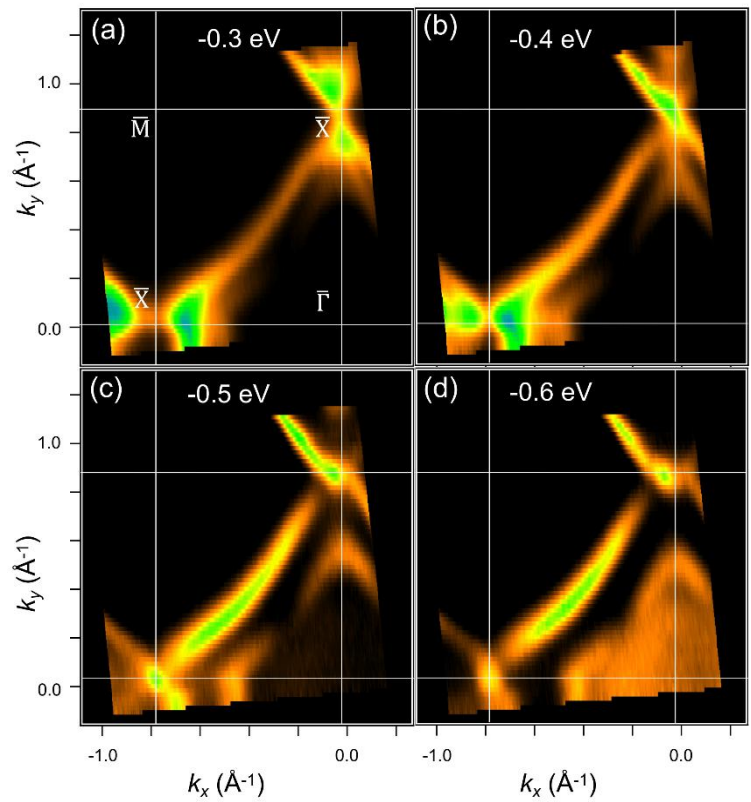

**Figure a4.** ARPES intensity maps of $\bar{\Gamma} - \bar{M} - \bar{X}$ plane at $E - E_F$ = -0.3 eV, -0.4 eV, -0.5 eV, and -0.6 eV with $h\nu$ = 21.2 eV.

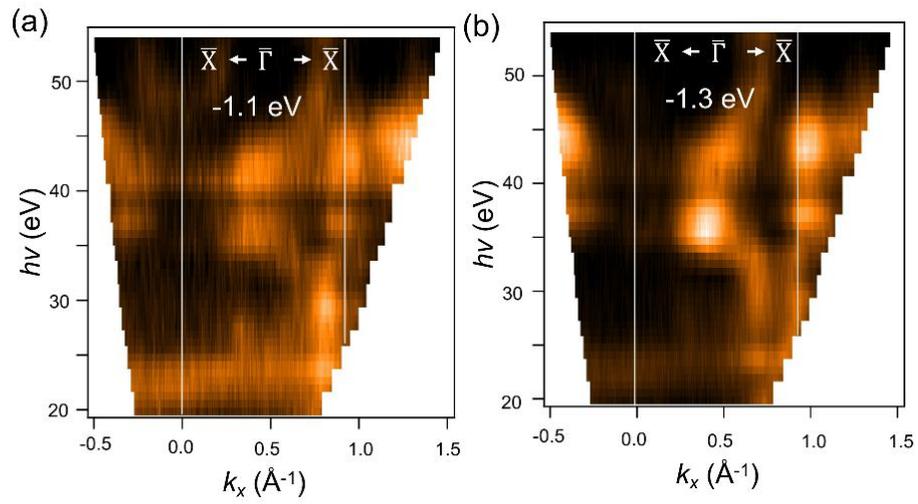

**Figure a5.** ARPES intensity maps for $E - E_F < -1$ eV along $\overline{X} - \overline{R}$ measured with $h\nu$ varying from 20 eV to 54 eV.